# Exoplanet *Terra Incognita*

Svetlana V. Berdyugina[1,2], Jeff R. Kuhn[2,3], Ruslan Belikov[4], Slava G. Turyshev[5]

[1]Kiepenheuer Institut für Sonnenphysik, Freiburg, Germany; [2]PLANETS Foundation, Kula, HI, USA; Institute for Astronomy, University of Hawaii, Pukalani, HI, USA; [3]Space Science and Astrobiology, NASA Ames, CA, USA; [4]Jet Propulsion Laboratory, Pasadena, CA, USA

**Abstract.** Exoplanet surface imaging, cartography and the search for exolife are the next frontiers of planetology and astrophysics. Here we present an overview of ideas and techniques to resolve albedo features on exoplanetary surfaces. Albedo maps obtained in various spectral bands (similar to true-colour images) may reveal exoplanet terrains, geological history, life colonies, and even artificial mega-structures or planet-engineering projects of advanced civilizations.

**Keywords.** Exoplanets, surface imaging, exolife, biosignatures, technosignatures, civilizations

## 1. Introduction

Extrasolar life (exolifie) detection by SETI techniques is ongoing for decades (Tarter et al. 2010), while exoplanet habitability and biosignatures are recent research fields (e.g., Grenfell 2017, Kiang et al. 2018, Catling et al. 2018). Here we explore how unambiguous exolife detection will be improved with direct or indirect imaging techniques that are capable of resolving the surfaces of exoplanets. The Earth seen as an unresolved planet shows a very marginal spectral signature of the chlorophyll red edge (a sign of the dominant photosynthetic life on present-day Earth) as well as oxygen, water and methane in the atmosphere (e.g., Sterzik et al. 2012). Modelling signals from distant Earth-like planets indicates that life-forms should cover more than 50% of the visible planet surface to be detected spectrally from unresolved planet (Berdyugina et al. 2016). In spatially resolved Earth's albedo maps, however, one can clearly see terrestrial global life colonies, such as vegetation (on a subcontinent scale), and our civilization footprints (at a higher resolution). Remote sensing of resolved surface structures on Earth can unambiguously identify their nature (e.g., Greenberg et al. 1993).

Here we discuss a similar strategy applied to exoplanets. Atmospheric biosignatures, such as the presence of biogenic gases and their disequilibrium (Krissansen-Totton et al. 2018), will help to identify exoplanets which possibly host life (perhaps similar to ours, since we focus our search on planets in a Water-based Habitable Zone, WHZ, while other HZ can be defined for non-Earth-like life based on different solvents and for different temperature ranges). There is however a risk of false-positive detections, since these gases may also form due to geochemical and photochemical processes affecting the composition of the planetary atmosphere (e.g., Meadows 2017). Therefore, obtaining resolved albedo maps of such Potentially Inhabited Exoplanets (PIE) will be crucial for identifying sources of biogenic gases and studying their distribution and evolution. In fact, both life and planetary activities are of great interest for planetology, astrophysics and astrobiology, as we will expand the realm of planetary and life morphology, which is now still limited to the Solar system.

While much of the current efforts in exoplanetary science are focused on identifying PIEs using atmospheric biosignatures, here we review emerging efforts to develop techniques and technologies which will provide resolved maps of exoplanets and open the door toward unambiguous detection of exolife. Some models for albedo maps and orbital parameters based on reflected flux or polarization light curves were demonstrated for exoplanets (e.g., Cowan et al. 2009; Fluri & Berdyugina 2010; Kawahara & Fujii 2011; Fujii & Kawahara 2012; Schwartz et al. 2016). More recently, an independent light-curve inversion formalism ExoPlanet Surface Imaging (EPSI) elaborated applications for planets with and without clouds, with seasonal variations, photosynthetic organism colonies and artificial structures built by advanced civilizations (Berdyugina & Kuhn 2017, hereafter BK17). It was demonstrated that inversions of multi-wavelength observations of such planets may enable detection of primitive and even advanced exolife with high confidence.

Here we distinguish between two approaches: *direct* and *indirect* Exoplanet Surface Imaging (EPSI). Direct EPSI requires km to Mm-size telescopes or interferometers on or near Earth, or alternatively targeted space missions beyond the Solar system. Such technologies are being developed and may be implemented in several decades if optimistic hopes for continued exponential technology growth are realized. Indirect EPSI is significantly less costly and in some cases comparably powerful: it can be done from the ground with 20m to 100m telescopes optimized for low scattered light. The technology to build such telescopes is within reach, such that investments now could yield an exolife telescope within a decade (Kuhn et al. 2018).

## 2. Indirect Exoplanet Surface Imaging

### 2.1. Cartography of unresolved Solar system bodies

Early cartography of some of the distant and small bodies in the Solar system, before space missions delivered their high-resolution images, was based on indirect imaging techniques. Historically important methods include occultations, limb, shadow and terminator measurements, as well as inversion of light curves measured at different rotational and orbital phases (see other reviews in this volume).

A light-curve analysis to reveal surface features of unresolved moons and asteroids was considered more than a hundred years ago (Russell 1906). Much later, such light-curve inversions have revealed the first surface maps of Iapetus (Morrison et al. 1975) and Pluto and Charon (see review by Buie et al. 1997). This allowed a study of their surfaces well before we saw images taken by spacecraft, which confirmed the main albedo features on planet surfaces recovered indirectly. Studying asteroid shapes and morphology from light-curve inversion (including polarimetry) is currently an active research field, which is enhanced by high-precision photometry from ground and space telescopes (e.g., Kaasalainen et al. 1992; Carbognani et al. 2012).

### 2.2. Gas Giant Exoplanets

The first light-curves of exoplanets were obtained for hot Jupiters, since they are large and very close to the host star, so they are not spatially resolved from stars. Infrared light-curves of hot Jupiters were obtained with the *Spitzer* space telescope (e.g., Harrington et al. 2006; Knutson 2007; Crossfield et al. 2010). Optical light-curves were obtained using ground-based polarimetry (Berdyugina et al. 2008, 2011) and space telescopes COROT (e.g., Snellen et al. 2009) and *Kepler* (Demory et al. 2011). From 2018, high-precision photometry of stars with exoplanets will be delivered by TESS in the optical. From 2020, JWST will provide measurements in the infrared. A combined analysis of optical and infrared light-curves provides stronger constraints on their atmospheres.

Hot Jupiters may be tidally locked to their orbital motion, with the rotational axis perpendicular to the orbital plane. In this case, their light-curves can resolve features only in longitude (no resolution in latitude). Infrared light-curves of hot Jupiters often show maximum radiation at orbital phases away from the exterior conjunction phase, which is when the planet is farther from the observer than the star. This was interpreted as the presence of hot spots and heat transport to the "dark side" of the planet (e.g., Harrington et al. 2006; Knutson 2007; Cowan & Agol 2008). Optical light curves provide a disk-average geometrical albedo of the planetary atmosphere, its dominant color, and may indicate the presence of clouds or hazes (Berdyugina et al. 2011; Evans et al. 2013; Pont et al. 2013).

Jupiter-, Saturn-, and Neptune-like gas giant planets at wider orbits (further from the host star) may rotate significantly faster than the orbital motion and be inclined to the orbital plane. The terminator misaligned with the rotational axis breaks the symmetry and allows spatial resolution in both longitude and latitude, when reflected light is sampled during several rotational periods at various orbital phases (preferably during the entire orbit). Since each flux measurement contains information on different parts of the partially illuminated planetary disk, one can employ a numerical inversion technique to "invert" such a 1D time series of flux measurements (light-curve) into a time-average 2D surface map (BK17). For gas giants, this helps resolving global circulation patterns imprinted into the cloud structure. Even when the planet is "rolling" on its orbit (like Uranus), there is still useful information to map its surface. This was demonstrated using Solar system Jupiter and Neptune images as inputs for EPSI inversions (BK17). Their Great Red and Great Dark Spots as well as zones and belts on Jupiter were successfully reconstructed (Fig. 1). This holds promise for studying global circulation on gas giant exoplanets on wider orbits in other planetary systems, which may be observed already now with large telescopes.

Inversions of thermal emission light-curves is similar to stellar and brown-dwarf light-curve inversions to recover brightness inhomogeneities due to magnetic spots or clouds (e.g., Berdyugina et al. 2002; Crossfield et al. 2014).

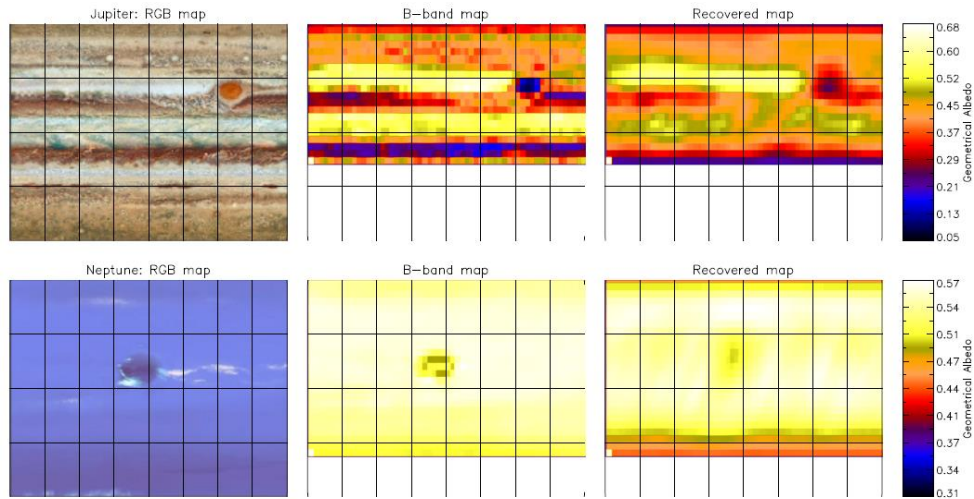

**Figure 1.** The EPSI inversion applied to the Solar system Jupiter and Neptune images as inputs (BK17). In the first column, are original RGB maps of higher resolution, in the second column are original albedo maps rebinned to a coarser grid in a particular band, and in the third column are recovered albedo maps in the same band.

## 2.3. Rocky Exoplanets

Rocky planets may have a variety of interesting surface features, if not fully covered by clouds, such as Venus. If they are covered in clouds, they would then be similar to the case of gas giant planets discussed in Section 2.2. In the Solar system, we see oceans, continents, deserts and forests on the Earth, ice caps, canyons and fluvial networks on Mars, volcanos on Io, heavily cratered terrains on the Moon, Mercury and moons of gas giants, glaciers on Pluto, and so on. Mapping albedo of such geological features on rocky exoplanets would provide hints to their geological history and internal structure and composition. Obviously, only large scale and high-contrast features can be distinguished and analysed.

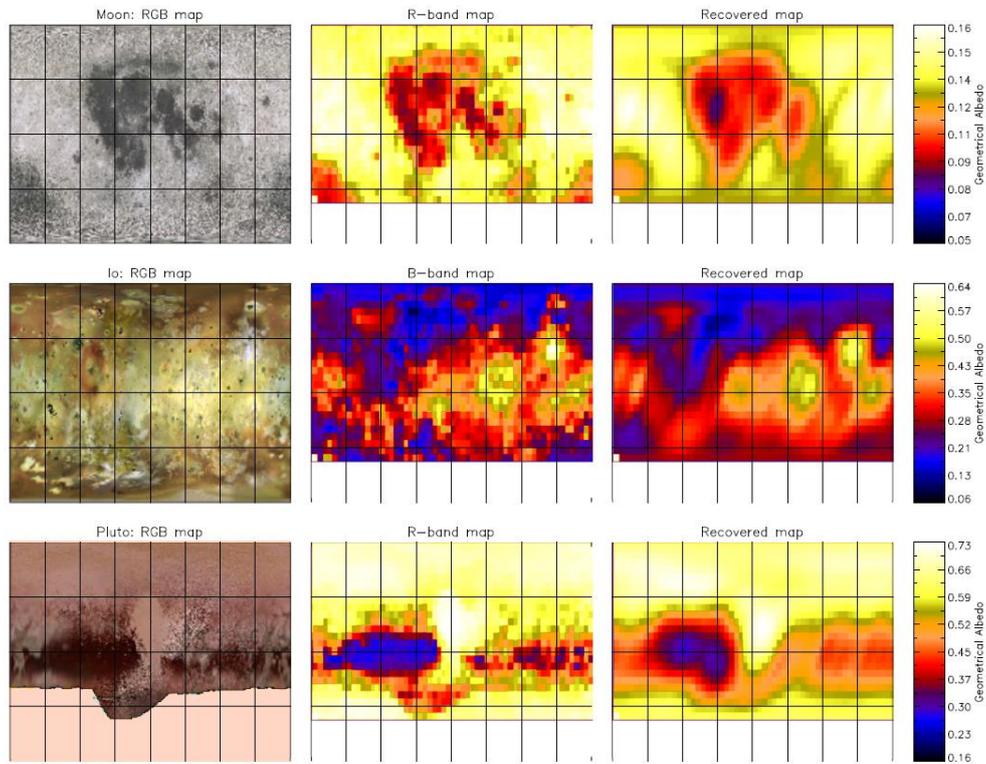

**Figure 2.** Same as Fig. 1 but for the Solar system rocky planetary bodies (BK17).

To resolve (indirectly) such surface features on Earth-size exoplanets requires high-precision light-curves (signal-to-noise ratio SNR>20) sampled at a range of orbital and rotational phases, especially if planets are intermittently covered by clouds, like the Earth. The interpretation of recovered albedo features in terms of geological, biological or artificial structures becomes possible. Here, EPSI techniques applied to the Solar system planet maps as models for rocky exoplanets can help in identifying such structures.

In Fig. 2 inversions for several solar planetary bodies are shown as examples. They remarkably recover large-scale terrains, continents, volcanoes (plums) and polar caps. Albedo maps obtained in various spectral bands can provide information on chemical composition of terrains (see Section 2.4).

## 2.4. Biosignatures

Detecting biosignatures and life on exo-Earths requires an analysis of the spectral content of the reflected, emitted or/and absorbed flux from an exoplanet. Light-curves measured in several spectral bands (e.g., blue, green, red, infrared) can provide color maps of exoplanets using the EPSI technique. This is a powerful tool for detecting living organisms on the surface, in particular photosynthetic organisms which have dominated the Earth for billions years.

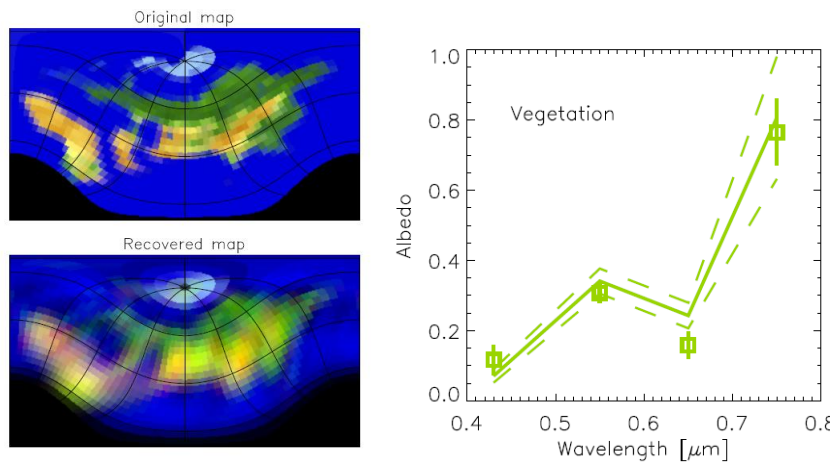

**Figure 3.** A test inversion for an exoplanet with Earth's surface samples in "true"-colour (left) and an average spectrum (right) of a small green area at mid-latitudes in the eastern part of the planet image with the red edge chlorophyll signature near 700 nm. "Measurements" from the recovered image in four broad spectral passbands are shown with symbols, while original spectra are shown with lines (BK17).

A test inversion for an exoplanet with the Earth's surface features, i.e., ice polar caps, ocean, deserts, and forest, is shown in Fig. 3 (BK17). Now, by resolving surface features, spectra of various areas can be extracted and explored for life signatures. An example average spectrum (in four broad spectral passbands) of a small green area at mid-latitudes in the eastern part of the planet reproduces well the red edge chlorophyll signature near 700 nm: a deep absorption in the visible wavelengths and high albedo in the near-infrared. Since photosynthetic organisms have a variety of biopigments absorbing light in broad bands, it may possible to distinguish them using such low-resolution spectra if they cluster in relatively large areas on the planetary surface. Also, measuring polarized light curves helps to identify biopigments, minerals and water reservoirs (Berdyugina et al. 2016). In particular, polarized light can measure glint from bodies of water (or other liquids, in general), which may be a good way to detect oceans (Karalidi & Stam 2012). Thus, spatially resolved images of exoplanets dramatically increases our chances to detect exolife, because we can carry out a spectral analysis of areas with high concentrations of living organisms (BK17).

## 2.5. Technosignatures

Although this is at present rather speculative, the EPSI technique allows also for detecting artificial mega-structures (AMS) constructed by advanced civilizations either on the surface or in the near-space of an exoplanet (circumplanetary space) (BK17). AMS could be of some regular shape and/or homogeneous albedo, or even above the surface as "geostationary" technology (e.g., for communications).

Low-albedo installations similar to our photovoltaic systems can be on the planet surface and in space. High-albedo installations, also on the surface or in the near-space, can redirect the incident stellar light, e.g., for heat mitigation by reflecting the light back to space. Such AMS may also efficiently absorb/reflect only a particular part of the spectrum, similar to photosynthetic organisms having specific spectral edges. An inversion example for a low-albedo installation in space (above clouds, Fig. 4).

These models show that recognizing AMS in inferred images requires large areas covered by them. Also, higher contrast structures with respect to the natural environment are easier to recover. Also, analyzing the spectral content of the reflected light can be useful.

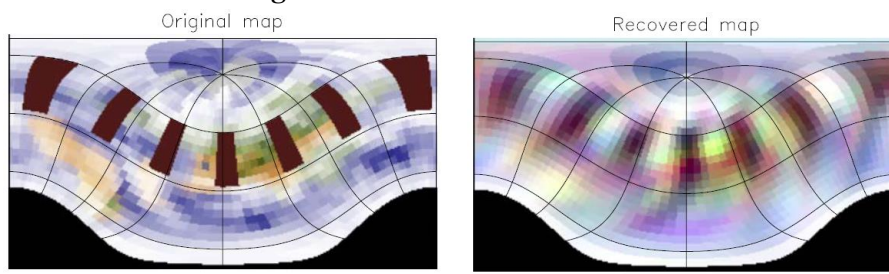

**Figure 4.** A test inversion for an exoplanet with artificial structures in "true"-colour (BK17).

## 2.6. Targets and telescopes for indirect EPSI

The Alpha Centauri system stars A, B, and Proxima Centauri are the closest stars to the Sun (4.3 ly). The only exoplanet discovered so far in this system is Proxima b (Anglada-Escudé et al. 2016), possibly a rocky planet in the WHZ of the M5 red dwarf. Other nearby possibly rocky planets in a WHZ are Ross 128 b (11 ly) orbiting an M4 red dwarf (Bonfils et al. 2017) and Tau Ceti planets e and f orbiting an orange dwarf G8 V (12 ly) (Feng et al. 2017). These nearby exoplanets are excellent candidates for first-time exoplanet surface imaging. In addition, more planets are statistically expected to exist around Alpha Centauri and other nearby stars (Burke et al. 2015).

Before the indirect EPSI technique can be applied, one has to first directly image an exoplanet, i.e., obtain an image with the star and planet sufficiently separated, so that the planet flux can be measured with sufficient signal to noise against the background of scattered light from the star and other background light sources. This is a challenge because of 1) a high star-to-planet contrast ratio and 2) small star-planet angular separation, especially for Earth-size planets. These challenges imply stringent constraints on the telescope size and performance.

The telescope must be large enough to collect enough photons from the faint exoplanet to make it detectable. We consider the nearest exoplanet Proxima b in an example calculation. The bright stellar background generally implies that the integration time to achieve a given signal-to-noise (SNR) ratio here scales more steeply than even the inverse fourth power of the telescope aperture diameter. In Fig. 5 (left panel) realistic SNR are computed for 1h exposure time in different passbands depending on the size of a telescope (BK17). Here the size of Proxima b is assumed 1.15 Earth radii, and its average geometrical albedo 0.2. This calculation employs also a "state-of-the-art" point spread function including scattered stellar light and speckles. We conclude that a low scattered light telescope with the effective diameter of ≥12m is necessary to obtain meaningful images of Proxima b. Since the assumed telescope is effectively diffraction limited, a space telescope would not significantly improve the Proxima b detectability (except in wavelengths where the telescope thermal emission or the atmosphere absorption dominate).

To achieve high-contrast measurements, one approach is to block the star with a coronagraph, starshade, and/or post-processing techniques, and directly image the planet, assuming it is spatially resolved from the blocked star. This has been achieved for stars with young, self-luminous, massive planets at orbits of tens of AU using such instruments as SPHERE at 8m VLT (e.g., Bonnefoy et al. 2016) and GPI on 8m Gemini (e.g., Ingraham et al. 2014). Currently deployed instruments are not yet capable of imaging WHZ exoplanets, but there are several concepts that may do it within the next 2 decades, and some as early as ~2 to 5 years from now. For example, imaging Alpha Centauri AB potential planets may be possible in the next 3-5 years in the visible light with a space telescope as small as 30cm, with a powerful enough coronagraph (Belikov et al. 2015; Bendek et al. 2015). Also, a 1.5-Earth radius WHZ planet around Alpha Centauri B may be within the capability of WFIRST using Multi-Star Wavefront Control (Sirbu et al. 2017), especially if post-processing methods can benefit from the greater brightness of Alpha Centauri.

Recently, a new concept for a ground-based high-contrast imaging system was proposed. The ExoLife Finder (ELF: Kuhn et al. 2018) is an array of off-axis telescopes with their primary mirrors being segments of the same parabola and served by individual adaptive secondary mirrors. It combines advantages of high-contrast off-axis optical telescopes with interferometric options to control wavefront. For instance, by changing phases of primary mirrors one can achieve nulling interferometry within a limited field of view (FOV), i.e., a low scattered light area, or dark spot, without a physical coronagraph. This spot can be moved within FOV for high-contrast imaging of an exoplanet at different orbital phases. In this respect, ELF's optical system is a hybrid interferometric telescope optimized for low scattered light and exoplanet observations.

A 20m ELF can continuously image at least a dozen of Earth-size planets in WHZ (Fig. 5, right panel). It will provide unique information on seasonal variations of their surfaces, atmospheres, and possibly life as well as possible geological activity (e.g., volcanoes). Its estimated cost is only about 100M$, and it can be built within 5 years. A 70m ELF (Kuhn et al. 2014; Moretto et al. 2016) could obtain images of all kinds of exoplanets within about 20pc, and its cost would be about 500M$.

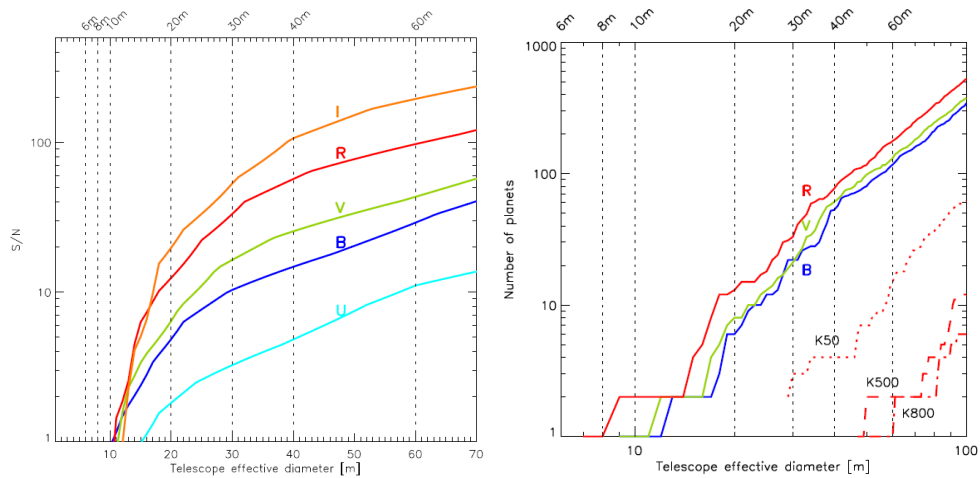

**Figure 5.** *Left:* Proxima b light-curve SNR depending on the size of the telescope, assuming 0.2 geometrical albedo and 1h exposure time. *Right:* Number of Earth-size exoplanets to be detected with the SNR≥5 depending on the size of the telescope, assuming one planet per nearby star, 0.2 geometrical albedo, and 4h exposure time. Solid lines are for high-contrast imaging telescopes (such as VLT and ELF), while dotted and dashed lines are for highly segmented Keck-style telescopes with a high scattered light background. (BK17).

Such large diffraction-limited optics will not soon be deployed in space because of extremely high cost (>10B$) and technological challenges. Note that

comparable aperture space and ground instruments can differ in cost by factors of 100 to 1000. Current space telescope concepts, such as NASA's LUVOIR or HabEx, may be implemented some time in the late 2030s, and they may be able to image same PIEs around nearby stars as a 20m ELF but in infrared bands which are not accessible from the ground. In the mid 2020s, also rather expensive (>1B$), ground-based Keck-style Extremely Large Telescopes (ELTs) promise to directly image exoplanets (Guyon et al. 2014), but it becomes recognized that additional significant efforts (e.g., high-dispersion coronagraphy, Wang et al. 2017) are needed to overcome large scattered light caused by their highly-segmented mirrors.

Yet another approach is to image planetary systems in the mid-infrared, where planet-to-star contrast ratio increases due to thermal radiation of the planet (e.g., Hinz et al. 2015). The ESO VLT 8m telescope instrumentation in the Southern hemisphere is being upgraded to employ this technique for imaging Alpha Centauri AB at 10 microns and search for exoplanets in this system. At this wavelength, we expect planet flux contributions from both reflection of the stellar light and own radiation of the planet. If planet radiation dominates, indirect EPSI will result in very low resolution maps (mainly resolved only in longitude). Also, to be sensitive to Earth-radius planets, thermal infrared imaging may require larger than 8m telescopes.

## 3. Direct Exoplanet Surface Imaging

### 3.1. Direct EPSI from the Earth distance

High-resolution direct surface imaging is the most expensive and most challenging for studying exoplanets. For example, a 2 km size optical telescope/interferometer is needed to theoretically resolve the disk of an Earth-size exoplanets (1 pixel image) at the distance of 1.3 pc (4.3 ly) from the Earth (Alpha Cen distance). Under the same conditions, a Jupiter-size planet can be resolved into an image of about 10×10 pixels, and a Sun-size star in a 100×100 pixel image. Correspondingly, to resolve directly the Proxima b planet into at least 10×10 pixel image requires a 20km-scale imaging system (possibly an array of smaller telescopes). This is a factor of 1000 larger than the ELF telescope which can provide indirect images of Proxima b of higher surface resolution (BK17). Also, building a 20km optical system is truly a technological challenge, and its cost may be forbidding.

### 3.2. Direct EPSI with the Solar Gravitational Lens Telescope

The 1.3 pc to the nearest exoplanet is about 270,000 AU. At about 550 AU from the Sun, a focal line of the solar gravitational lens (SGL) begins, i.e., the solar gravitational field can focus light from a faint, distant source along that line by bending photon trajectories (Fig. 6). The remarkable optical properties of the SGL include major brightness amplification (~$10^{11}$ at the near-infrared wavelength of $\lambda = 1$ μm) and extreme angular resolution (~$10^{-10}$ arcsec) within a narrow field of view (Turyshev & Toth, 2017). It was proposed to position a telescope beyond 550 AU and use the SGL to magnify light from distant objects on the opposite side of the Sun, e.g., Proxima b (Eshleman 1979; Maccone 2009; Turyshev & Toth 2017, Turyshev et al. 2018).

The angular diameter of an Earth-like exoplanet at 30 pc is $1.4 \times 10^{-11}$ rad. To resolve the disk of this planet as a single pixel, a telescope array with a baseline of ~74.6 km would be needed. Resolving the planet with $10^3$ linear pixels would require a baseline ~$1 \times 10^5$ km (~$12 R_\oplus$), which is not feasible. In contrast, a 1-m telescope, placed at the focal line of the SGL at 750 AU from the Sun, has a collecting area equivalent to a diffraction-limited telescope with diameter of ~80 km and the angular resolution of an optical interferometer with a baseline of $12 R_\oplus$. While building an instument with the SGL's capabilities is far beyond our technological reach, we can use the SGL's unique capabilities to capture megapixel images of exoplanets.

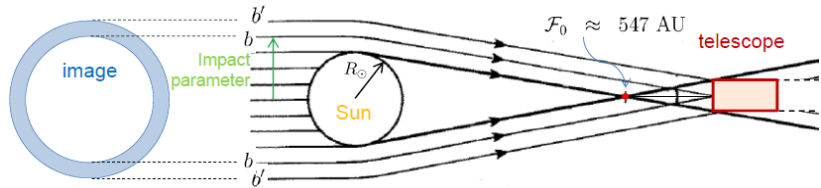

**Figure 6.** Imaging of an exo-Earth with solar gravitational Lens. The exo-Earth occupies about 1km×1km area at the image plane. Using a 1m telescope as a 1 pixel detector provides a 1000×1000 pixel image.

A modest telescope at the SGL could be used for direct imaging of an exoplanet. The entire image of an exo-Earth is projected by the SGL into an instantaneous cylindrical volume with a diameter of ~1.3 km. Moving outwards while staying within this volume, the telescope will take photometric data of the Einstein ring around the Sun, formed by the light from the exoplanet. The collected data will be processed to reconstruct the desirable high-resolution image by relying on modern deconvolution techniques.

To sample a planet image within the Einstein ring, an SGL telescope will have to operate on a pixel-by-pixel basis by moving the spacecraft in the image plane. A $10^3 \times 10^3$ pixel image will therefore require $10^6$ pointings with the spacecraft step of ~1m. Since each pointing corresponds to a different impact parameter (see Fig. 6) with respect to the Sun, one has to account for light from adjacent surface areas on the planet. Also, to place and keep the telescope on the focal line, high precision astrometry of the planetary orbit is needed, as the telescope spacecraft will have to compensate stellar and planetary motion. The parent star will be completely resolved from the planet, and its light will be amplified many AU away from the optical axis, making the parent star contamination issue negligible. For example, an exo-Earth at 30pc can be imaged by a 1m SGL telescope operating at 650AU with SNR of 7 in ~1 week of integration time, when accounting for zodi background, the solar corona brightness, coronagraphic suppression of $10^{-6}$, spacecraft jitter, and other realistic losses (Turyshev & Toth 2017, Turyshev et al., 2018).

The SGL telescope is a single-target instrument and, therefore, a powerful follow up mission for a target with signatures of life revealed by EPSI mapping and spectral analysis. Given the recent progress in finding exoplanets and in their characterization with indirect techniques (Section 2), it is a matter of a relatively short time that we will have plenty of exciting targets to explore with the SGL. It would take however ~30 years of "cruise" before the spacecraft would reach 600AU with the current technology yielding escape velocity of 15-20 AU/yr, see (Turyshev et al., 2018). In comparison, the farthest man-made object Voyager 1 moves at the speed of 3.6 AU/yr (17 km/s), and after 40 years it is at 141 AU from the Sun. A mission to the SGL enables the unique possibility of direct high-resolution imaging and spectroscopy of a habitable Earth-like exoplanet.

### 3.3. Direct EPSI with a flyby probe

The Alpha Centauri system may in some distant future also be explored by interstellar probes. The 4.3 ly distance implies a 43 year mission, if a tiny probe can be accelerated to 10% speed of light (~30,000 km/s, or 6300 AU/yr). In comparison, the largest geocentric velocity achieved by the DLR/NASA Helios 1 spacecraft was ~100 km/s, and the largest heliocentric velocity by the NASA Juno mission was ~60 km/s. Thus, achieving 10% speed of light is yet quite challenging, but significantly reducing the size of the probe may help.

It was recently proposed to develop a light-sail spacecraft of only 1g or less to fly to the Alpha Centauri system at 20% of the speed of light (Lubin 2016). The spacecraft *StarChip* is being developed by the Starshot project of Breakthrough Initiatives and to be propelled by a 100 GW square-kilometer laser array. A thousand units would be released to compensate for possible interplanetary and interstellar dust collisions. Planet images will be obtained during a flyby with a high speed, so deconvolution will be still needed to achieve higher spatial resolution, but a slow-down maneuver seems possible (Heller & Hippke 2017; Heller et al. 2017). It is desirable that onboard miniature cameras can deliver multi-passband images for a spectral, polarimetric and morphological analysis, so that the chemical, structural and hopefully biological composition of these planets can be studied directly. Another interesting opportunity is to arrange a direct probe of the planetary atmosphere and even surface, if some of the probes can intercept the planet.

## 4. Conclusions

In just a few years we can discover *terra incognita* on exoplanets and start their exploration in terms of surface composition, morphology, diversity, frequency, life/civilization detection, etc. Similar to the last century indirect imaging of small and distant Solar system bodies, we can start practicing exoplanetology and exogeology, and even exobiology and exosociology of extraterrestrial life and civilizations.

Albedo maps in optical and infrared bands can be obtained using light-curve inversions for a dozen of Earth-size planets with a 20m telescope like ExoLife Finder and for almost all exoplanets within about 20 pc with a 70m ELF. These telescopes are optimized for low scattered light and exoplanet observations. The technology to build such telescopes is within reach, such that they could become operational within a decade.

In a more distant future, concurring space outside the Solar system and sending probes to selected exoplanets could be possible. Such missions are technology drivers and with the help of Moore's law may become reality in a couple of decades. Meanwhile we should get busy with the indirect EPSI in order to select the most promising targets for direct probes.